\documentclass[a4paper]{article}

\usepackage{INTERSPEECH2016}

\usepackage{amsmath}
\usepackage{epsfig}
\usepackage{times}

\title{Text-Independent Speaker Verification Based on Deep Neural Networks and Segmental Dynamic Time Warping}

\makeatletter
\def\name#1{\gdef\@name{#1\\}}
\makeatother
\name{{\em  Mohamed Adel$^1$, Mohamed Afify$^1$ and Akram Gaballah$^2$}
}
\address{$^1$Microsoft Advanced Technology Lab, Cairo, Egypt \\
$^2$Microsoft Corporation, Redmond, WA, USA \\
  {\small \tt \{a-moadel, mafify, akrgab\}@microsoft.com}
}

\begin{document}
\maketitle

\begin{abstract}
In this paper we present a new method for text-independent speaker verification that combines segmental dynamic time warping (SDTW)
and the d-vector approach. The d-vectors, generated from a feed forward deep neural network trained to distinguish between speakers, are used as features to 
perform alignment and hence calculate the overall distance between the enrolment and test utterances.
We present results on the NIST 2008 data set for speaker verification where the proposed method outperforms the conventional i-vector baseline with PLDA scores and outperforms d-vector approach with local distances based on cosine and PLDA scores. Also score combination with the i-vector/PLDA baseline leads to significant gains over both methods.
\end{abstract}

\section{Introduction}
\label{introduction}
Speaker verification is the process of confirming whether an input utterance belongs to a claimed speaker.There are many popular 
approaches to the problem including Gaussian mixture model (GMM)  \cite{gmm}, i-vector \cite{ivector} and more recently deep learning \cite{dvector}. 
Speaker verification could be further classified into text-dependent 
and text-independent. In the text-dependent mode, both the enrolment and test utterances have the same text, while in the text-independent case the user can enroll and test with any text.

The d-vector approach \cite{dvector} has been originally proposed for text-dependent speaker verification. The basic idea is to train a deep neural network to
learn a mapping from the spectral input to the speaker identity. An intermediate layer (embedding) is then extracted for each input frame. The extracted embedding is averaged over the input utterance and used as a speaker representation, called the d-vector, similar to the i-vector. The d-vector is then used for speaker verification by applying a cosine distance or probabilistic linear discriminant analysis (PLDA) \cite{PLDA}.

Several variants are proposed to improve on the original d-vector idea or generalize it to the text-independent
scenario. An end-to-end loss, more related to speaker verification, is proposed in \cite{dvectore2e} and applied to train feed-forward and LSTM architectures. 
The latter end-to-end loss is generalized in \cite{dvectorge2e} and applied to both text-dependent and text-independent
verification. Different attention 
mechanisms proposed in \cite{shixiong2016,dvectorattn} yield improvement over simple averaging 
for calculating the d-vector. Deep speaker \cite{deepspeaker} uses different architectures, including convolutional networks with residual connections
and gated recurrent unit (GRU) networks, and train them using the triplet loss. The proposed architectures show good results for both text-independent
and text-dependent speaker recognition. Interestingly, it is also shown in \cite{deepspeaker} that a network trained for text-independent verification can be adapted using
task-dependent data. This is important because the d-vector does not work very well for small task-dependent data size \cite{dvectorkenny}.
 In \cite{dvectorti}, the intermediate representation from an LSTM, trained either separately 
or jointly with speech recognition, is used for text-independent 
speaker verification on Hub5 and Wall Street Journal (WSJ) data. The 
work of \cite{dvectore2eti} trains a neural network with temporal pooling in an end-to-end fashion for text-independent 
speaker verification, similar in spirit to \cite{dvectore2e}, 
and present results on telephone speech with various durations. A time delay neural network (TDNN) is trained using cross entropy and the resulting embedding is used 
for PLDA scoring in \cite{dvectorpldati}. The results are presented on NIST speaker verification tasks for various test durations. A major finding is that 
the d-vector outperforms the conventional i-vector for short duration segments while the latter is better for longer duration. The latter work is recently extended by data
augmentation and applied to various data sets in \cite{x-vectors}. It is also worth mentioning
approaches inspired by i-vector and PLDA where the whole i-vector/PLDA system is formulated and trained as a network \cite{ivectorpldae2e}.

Instead of averaging the d-vectors over the whole utterance, as is typically done in conventional approaches, 
we propose to keep the sequences of d-vectors of the enrolment and test utterances.
We then align the two sequences to come up with an accumulated score for text-independent speaker verification. In \cite{dvectordtw}, dynamic time warping (DTW) \cite{dtw} is used to find the best 
alignment and hence the minimum distance between two sequences of d-vectors for text-dependent verification. However, conventional DTW with path constraints 
will not lead to meaningful alignments in the text-independent case. This is because path constraints might be too restrictive to find the potentially non-monotonic alignments. Here we use segmental dynamic time warping (SDTW) to align the resulting two sequences of d-vectors and experiment with both cosine distance and 
PLDA for measuring the local distance between pairs of d-vectors. We present results on the NIST 2008 speaker verification task.

Segmental DTW has been proposed for automatic pattern discovery of speech in \cite{segdtw}. It was then applied to keyword 
spotting \cite{kwssegdtw} and speaker segmentation \cite{spksegdtw}. In this article, we combine the d-vector
with SDTW to do text-independent speaker verification. At a high level, SDTW finds multiple partial paths of two utterances and hence 
could discover parts of the utterances that exhibit certain similarities.We combine the 
scores of these paths to come up with a similarity score between the two utterances and use it for verification.

The rest of the paper is organized as follows.Section 
\ref{SDTW} briefly describes SDTW. Speaker verification using the d-vector and SDTW is described in detail in Section \ref{algorithm}.Finally, 
experimental results on NIST 2008 and conclusion are given in Sections \ref{experiments} and \ref{conclusion} respectively.

\section{Segmental Dynamic Time Warping}
\label{SDTW}
In this section we briefly describe segmental DTW. In its basic form \cite{dtw}, DTW finds the optimal
global alignment and the accumulated distance between two sequences $\cal X$ and $\cal Y$. 

Assume ${\cal X} = (x_{1},x_{2},....x_{I})$ and
${\cal Y} = (y_{1},y_{2},...y_{J})$, an alignment $\phi$ is given by
\begin{equation}
\phi = (i_{k},j_{k})  \;\;\;\;\;   k=1,......T   \label{alignment}
\end{equation}
where $i_{k}$ and $j_{k}$ are indices from the two sequences and $T$ is the alignment length such that $i_{T}=I$ and $j_{T}=J$. The associated accumulated distance is given by
\begin{equation}
D_{\phi}({\cal X}, {\cal Y}) =  \sum_{k=1}^{T} d(x_{i_{k}},y_{j_{k}})    \label{accumulated}
\end{equation}
where $d()$ is a local distance. In this work we use distance measures based on the cosine similarity and probabilistic linear discriminant analysis (PLDA). 

\begin{figure}[h!]
  \includegraphics[width=\linewidth]{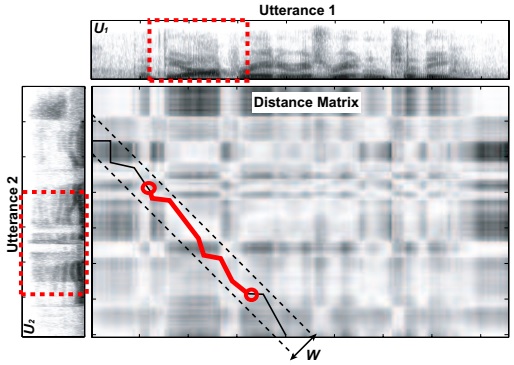}
  \caption{Segmental dynamic time warping \cite{segdtw}.Two utterances, Utterance 1 and Utterance 2, are shown. The central rectangle, labelled distance matrix,shows the local distances between the vectors of the two utterances. The diagonal region of width $W$ (dependent on $R$) represents performing DTW between the dashed red boxes of the two utterances. The red path is the best fragment of length at least $L$ of the optimal path.} 
  \label{fig0}
\end{figure}

Given a distance measure and a set
of constraints, DTW calculates the optimal path and the associated accumulated distance using dynamic programming \cite{dtw}.
The set of constraints are important to obtain a physically plausible alignment. The so-called adjustment window condition ($|i_{k}-j_{k}| \leq R$) \cite{dtw}
ensures the aligned indices of the two sequences are not very far apart. If the
two sequences grossly violate the constraints, DTW will most likely fail to find a 
good alignment.

Segmental DTW generalizes DTW by finding a set of partial alignments between two sequences.By allowing partial alignments, SDTW can potentially
find well matched sub-sequences even if the two sequences are not fully matched.
Formally, with a constraint parameter $R$ and
utterances lengths $I$ and $J$ respectively, we obtain multiple alignments by running DTW on regions that start at:
\begin{eqnarray}
((2R+1)k+1,1),  &   0 \leq k \leq \lfloor\frac{I-1}{2R+1}\rfloor  \nonumber \\
(1,(2R+1)l+1),  &   1 \leq l \leq \lfloor\frac{J-1}{2R+1}\rfloor.  \label{coordinates}
\end{eqnarray} 
Each region will be limited to a diagonal region depending on $R$ and hence represents a partial alignment of the two
sequences. One region is shown in Figure \ref{fig0} taken from \cite{segdtw}. $W$ in the figure refers to the width of the region where a partial alignment is calculated. This is determined by the constraint parameter $R$. We refer to these partial alignments as $\phi_{r}$ where $r=1,......,N_{R}$. Now, given a length constraint parameter $L$, we
find for each local alignment path, the fragment of length at least $L$, shown in red in the figure, that has the minimum average distortion. This can be efficiently
obtained as outlined in \cite{segdtw} and references therein.  

To summarize, given parameters $R$ and $L$, SDTW of two sequences gives a set of fragments of length at least $L$ and their associated
scores as $(\phi_{r},s_{r})$ for $r=1,.......N_{R}$. We will show in the next section how to use these partial scores for speaker
verification. The best parameters $R$ and $L$  are determined empirically.

\section{Speaker Verification Using d-vector and SDTW}
\label{algorithm}
In this section we will describe how to combine d-vector and SDTW for text-independent speaker verification. 
Also the network architecture and training used to generate d-vectors will be described in Section \ref{dvectorarch}. We will
focus on the case of single enrolment and single test.The generalization to multiple enrolments and multiple 
tests is straightforward.

In speaker verification, the distance between the enrolment and test utterances is calculated and compared to 
a threshold to either accept or reject the claim. For the text-independent case both enrolment and test have
different phonetic content. As discussed above, SDTW between two sequences provides a set of partial alignments
and their distances. The distance between the sequences can be obtained by combining the distances of the partial alignments.
In preliminary experiments, we tried using the average, the average of the lowest-K and the minimum with very similar performance.
Thus, we will report the average in the rest of the work.Once the distance is obtained it is compared to a threshold for the
verification decision.

Motivated by recent success of deep learning techniques in speaker verification we apply SDTW at the d-vector level.We 
first generate a sequence of d-vectors for both the enrolment and test utterances then apply SDTW to the resulting d-vectors.We summarize the enrolment and verification
phases below.
\begin{enumerate}
\item Enrolment Phase
\begin{itemize}
\item Starting from enrolment utterance ${\cal X}_{e}$ create sequence of enrolment d-vectors
${\cal X}^{d}_{e}=x^{d}_{e,1}$  $,x^{d}_{e,2},......x^{d}_{e,N_{e}}$. 
This is obtained by running a fixed length window on the enrolment utterance and advancing it
by a fixed step. Each window is input to the network to produce the corresponding d-vector.
Please note that the time index of the enrolment sequence follows the input step and not the
frames. In the case we advance the window by one frame they will coincide.   
\end{itemize}
\item Verification Phase
\begin{itemize}
\item Starting from test utterance ${\cal X}_{t}$ create sequence of test d-vectors
${\cal X}^{d}_{t}= x^{d}_{t,1},x^{d}_{t,2},......x^{d}_{t,N_{t}}$ similar to the enrolment.
\item Run SDTW, with parameters $R$ and $L$, on the enrolment and test d-vector sequences ${\cal X}^{d}_{e}$ and  ${\cal X}^{d}_{t}$.This will result, as discussed
above, in a set of partial paths and their corresponding scores  $(\phi_{r},s_{r})$ for $r=1,.......N_{R}$.
\item Obtain the score of the test utterance by averaging the scores of the partial paths. This score is then compared
to a threshold to make the verification decision.   
\end{itemize}
\end{enumerate} 

\subsection{D-vector Network Architecture and Training}
\label{dvectorarch}
Any network architecture can be used with the proposed method. In this work we use a simple feed-forward architecture. 
The input dimension is 1386, as explained below, it consists of feature vector dimension of size 66
and context window of size 21. This is followed by 5 hidden layers that operate on the frame level of sizes 2048, 2048,
1024, 1024 and 512 respectively. All layers use ReLU non-linearity and batch normalization. This is followed by a temporal
pooling layer that operates on the input segment. Following temporal pooling is a hidden layer of size 128 that operates
on the segment level and also uses ReLU and batch normalization. Finally, there is the output layer that uses cross entropy
and softmax. The output layer corresponds to the 5000 speakers having the largest number of segments in the training data.
The d-vector is extracted from the last hidden layer, after temporal pooling, of size 128. Excluding the softmax layer, the
network has about 10M parameters.Dropout with keep parameter 0.75 is used after the second hidden layer which has about 4M parameters. 

Segments of length 200 frames with an advance of 50 frames are extracted from the training data. Data from the most frequent 5000 speakers are kept with about 3000 segments/speaker.
This leads to a total of about 15M segments of size 200 labelled with the corresponding speaker. We did some experiments on window duration selection or using random window size but found the selected size to work best. Training optimizes the cross entropy criterion. Other criteria as triplet loos could be used but these typically need CE initialization and could be tried in future work. The network is randomly initialized. Mini-batches of size 70 segments are randomly formed from the above segments and used to optimize the weights using SGD with momentum. The learning rate is reduced after every sweep through the data to prevent over-fitting.

\section{Experimental Results}
\label{experiments}
In this section we present experiments to verify the proposed method. We first present the training and testing data, followed by the baseline setup for i-vector/PLDA and d-vector and experimental results.

\subsection{Training Data and Testing Setup} 
The training data consists of about 4000 hours from the English Fisher and the NIST 2004,2005 and 2006 telephone corpora sampled at 8 kHz. Voice activity detection (VAD)using an energy-based criterion is applied to the data. 22 log filter-bank energies (LFB) are extracted together with their first and second derivatives leading to 66-dimensional feature vector that is 
used during training and testing. We use window of 21 frames, centred around the current frame, leading to a network 
input size of 1386 \footnote{We tried several window sizes and 
found that 21 is the best.}.The common evaluation condition of NIST 2008 SRE is used for testing. We use both telephone and interview data 
comprising all conditions C1-C8.

\subsection{Baseline System}
For comparison we use an i-vector/PLDA baseline and a d-vector baseline. The configuration of these systems are as follows:
\begin{itemize}
\item i-vector/PLDA: This follows the system in \cite{ivector} and based on our previous experiments we set the UBM size to 2048 and the
i-vector size to 400. We always test i-vector with PLDA as in \cite{PLDA}. The PLDA dimension is set to 200. After generating
i-vectors for the training data we project them to dimension 200 using LDA then apply centering, whitening and length normalization.
The PLDA is then trained on the transformed data. The same processing is done on the test data and the PLDA score is used for verification.
\item d-vector: The baseline d-vector system works as follows. First a sequence of d-vectors are generated by sliding a window over
the test utterance as described above. The resulting sequence of d-vectors is then averaged to yield a single vector representation
for the utterance. We use two configurations.  The first uses cosine distance for scoring while the second, similar to i-vector, uses PLDA. The d-vector size is 128 for both configurations.
\end{itemize}
\subsection{Results}
Table \ref{baseline-tab} shows the equal error rate (EER) averaged over the 8
conditions. The second column presents the results of the following systems: i-vector/PLDA, d-vector with cosine
scoring, d-vector/PLDA and d-vector with cosine and d-vector/PLDA with SDTW. The latter two use $R=1$ and $L=30$. 
The i-vector/PLDA EER in the second row is a reasonable baseline compared to other results obtained on NIST 2008. Although this can be further optimized using
gender-dependent models and phonetically-aware features, adding these can also
benefit the d-vector and hence are not tried here.
The d-vector with cosine scoring shows
significantly worse performance than the i-vector while the d-vector/PLDA is better than the baseline d-vector. The d-vector/PLDA is still worse than the i-vector/PLDA.
Results regarding the latter point are mixed. For example, \cite{deepspeaker} shows 
excellent results with only cosine scoring while \cite{dvectorpldati} shows only results
with PLDA. We believe that for public corpora where there are only few thousand speakers for training, training data mainly consists of telephone speech while test data has varying acoustic conditions, the network will not fully learn to normalize the acoustic
condition. Hence, PLDA will provide desirable normalization on top of d-vector and
can potentially lead to better results as shown here. In \cite{dvectorpldati} i-vector/PLDA is slightly better than d-vector/PLDA when testing with the full utterance\footnote{We test on the core condition where the average utterance length is around 2 minutes.}. Finally, we can see decent gains by using SDTW on top of d-vector both with cosine and PLDA. In particular, the d-vector/PLDA with SDTW performs better than i-vector/PLDA.
The third column shows the results of combining i-vector/PLDA, on the score
level with weight 0.5, with other systems. In all cases we see significant gains from the combination.

\begin{table}[ht]
\caption{\it{Baseline and SDTW EER results averaged over all 8 conditions of NIST 2008 core test. Also combined results with i-vector/PLDA are shown in the third column.}}
\vspace{2mm}
\centerline{
\begin{tabular}{|c|c|c|}
\hline
Method &  EER &  EER (+ i-vector)\\ \hline \hline
i-vector/PLDA & 7.15\% & NA \\ \hline
d-vector & 10.39\%  & 5.45\% \\ \hline
d-vector/PLDA& 8.46\%  & 4.23\%  \\ \hline
d-vector + SDTW &   8.17\% & 4.14\% \\ \hline
d-vector/PLDA + SDTW &   6.41\% & 3.95\%  \\ \hline
\end{tabular}
}
\label{baseline-tab}
\end{table}

\begin{figure}[h!]
  \includegraphics[width=\linewidth]{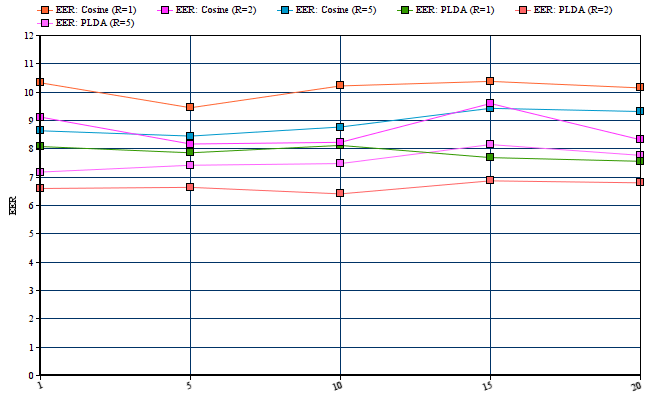}L
  \caption{The results of SDTW with and without PLDA for various 
  values of $R$ and $L$. The colored curves stand for cosine and PLDA for $R=$1,2 and 5. The horizontal axis stands for values of $L$ and the vertical axis for EER.}
  \label{fig1}
\end{figure}

Figure \ref{fig1} shows SDTW results for both cosine and PLDA scoring with varying $R$ and $L$. Six curves with different colors correspond to $R=1$, $R=2$, and $R=5$ and cosine and PLDA scoring. The horizontal axis stands for different values of $L$. It can be observed that PLDA results are significantly better than cosine results
for all values of both parameters. Generally speaking, we observe that small values of $R$ give better performance because we sample at a rather coarse rate of 50 frames. Also relatively large $L$ tends to give better result as longer segments carry meaningful speaker information. Also in the figure we observe a fairly stable region for selecting
the $R$ and $L$.

To gain more insight, we show in Table  \ref{individual-tab} the results of individual
conditions. These correspond to the second column of Table \ref{baseline-tab} before
averaging. It is clear that the d-vector approach is significantly better 
than i-vector for the telephone conditions (C6-C8) and significantly worse
for the microphone (interview) conditions (C1-C3). The d-vector is also better for
the mixed conditions (telephone/interview) C4 and C5. As the training data for the
network and PLDA consist mainly of telephone speech. We might argue that, for
the available amount of training data and network architecture, the network is
not able to fully normalize for the acoustic condition. PLDA helps for acoustic
normalization in almost all conditions. Also SDTW shows significant improvement
in almost all conditions. 



\begin{table*}[ht]
\caption{\it{Baseline and SDTW EER results on NIST 2008 conditions C1-C8.}}
\vspace{2mm}
\centerline{
\begin{tabular}{|c|c|c|c|c|c|c|c|c|}
\hline
Method & C1 & C2 & C3 & C4 & C5 & C6 & C7 & C8  \\ \hline \hline
i-vec/PLDA  &  11.07\% & 1.54\% & 11.26\% & 9.68\% & 9.00\% &	8.53\% &	11.47\% &	12.93\%   \\ \hline
d-vec & 20.87\% &	1.91\% &	20.92\% &	6.84\% &  7.36\% &	6.21\% &	8.72\% &	10.32\%  \\ \hline
d-vec + PLDA  & 17.06\% &	1.28\% &	15.93\% &	7.06\% & 4.39\% &	6.12\% &	6.74\% &	9.09\%   \\ \hline
d-vec + SDTW  &  19.23\% &	2.02\% &	19.49\% &    1.57\% & 3.61\% &	4.09\% &	5.71\% &	9.66\%  \\ \hline
d-vec PLDA + SDTW  &  14.06\% &	1.09\% &	14.69\% &	1.43\% & 4.64\% &		4.76\% &		3.78\% &		6.79\%   \\ \hline
\end{tabular}
}
\label{individual-tab}
\end{table*}

\section{Conclusion}
\label{conclusion}
We propose segmental DTW to align the d-vectors of the enrolment and test utterances for text-independent speaker verification.
Compared to the conventional d-vector, which averages the d-vectors over the whole utterance, alignment can potentially   
find better matching parts of the enrolment and test utterances and hence reduce bias due to phonetic content.
Compared to conventional DTW, Segmental DTW can find good partial alignments even if the two utterances are grossly mismatched. The proposed
method is tested on the core condition of NIST 2008 where the utterances are relatively long and shows improvement over the baseline d-vector with and without PLDA scoring. 
Combining with i-vector/PLDA provides interesting gains in all cases. Future work includes improving the d-vector itself by exploring more sophisticated architectures
recurrent and convolutional networks and using other training criteria as triplet loss.
We also plan to test on other corpora like NIST SRE 2010, SITW and VoxCeleb.

\section{Acknowledgement}
We thank Ahmed Ewais and Mohamed Yahia for working on an earlier setup for this task and Lana Chafik for help with some experiments.


\begin{thebibliography}{90}
\bibitem{gmm} D. Reynolds, T.F. Quatieri and R.B. Dunn, "Speaker Verification Using Adapted Gaussian Mixture Models," Digital Signal Processing, Vol. 10, 
pp. 19-41, 2000.
\bibitem{ivector} N. Dehak, P. Kenny, R. Dehak, P. Dumouchel, and P. Ouellet "Front-End Factor Analysis for Speaker
Verification," IEEE Transactions On Audio, Speech And Language Processing, Vol. 19, No. 4, May 2011.
\bibitem{dvector}E. Variani, X. Lei, E. McDermott, I. Lopez-Moreno and J. Gonzalez-Dominguez, "Deep Neural Networks for Small Footprint
text-dependent Speaker Verification," In Proc. ICASSP 2014.
\bibitem{dvectore2e} G. Heigold, I. Moreno, S. Bengio and N. Shazeer," End-to-End Text-Dependent Speaker Verification," In Proc. ICASSP 2016.
\bibitem{dvectorge2e} L. Wan, Q. Wang, A. Papir and I. Lopez Moreno, "Generalized End-to-End Loss for Speaker Verification," arXiv preprint arXiv:1710.10467, 2017.
\bibitem{dvectorattn} F. Chowdhury, Q. Wang, I. Lopez Moreno and L. Wan, "Attention-Based Models for Text-Dependent Speaker Verification," arXiv preprint arXiv:1710.10470, 2017.
\bibitem{deepspeaker} C. Li, X. Ma, B. Jiang, X. Li, X. Zhang, X. Liu, Y. Cao, A. Kannan and Z. Zhu, "Deep Speaker: An End-to-End Neural Speaker Embedding System," arXiv preprint arXiv:1705.02304, 2017.
\bibitem{dvectore2eti} D. Snyder, P. Ghahremani, D. Povey, D. Garcia-Romero, Y. Carmiel and S. Khudanpur, "Deep Neural Network Speaker Embeddings for end-to-end Speaker Verification," In Proc. SLT2016.
\bibitem{dvectorpldati} D. Snyder, D. Garcia-Romero, D. Povey and S. Khudanpur, "Deep Neural Network Embeddings for Text-Independent Speaker Verification," In Proc. Interspeech 2017.
\bibitem{ivectorpldae2e} J. Rohdin, A. Silnova, M. Diez, O. Plchot, P. Matejka and L. Burget, "End-to-End DNN Based Speaker Recognition Inspired by i-vector and PLDA," arXiv preprint arXiv:1710.02369, 2017.
\bibitem{dvectorkenny} G. Bhattacharya, J. Alam, T. Stafylakis and P. Kenny, "Deep Neural Network Based Text-Dependent Speaker Recognition: Preliminary Results,"
In Proc. Odyssey 2016.
\bibitem{dvectorti} Z. Tang, L. Li and D. Wang, "Multi-Task Recurrent Model for Speech and Speaker Recognition," arXiv preprint arXiv:1603.09643, 2016.
\bibitem{dvectordtw} L. Li, Y. Lin, Z. Zhang and D. Wang, "Improved Deep Speaker Feature Learning for Text-Dependent Speaker Verification," 
arXiv preprint arXiv:1506.08349, 2015.
\bibitem{ivectoradapt} V. Gupta, P. Kenny, P. Ouellet, and T. Stafylakis, "I-Vector-Based Speaker Adaptation Of Deep
Neural Networks For French Broadcast Audio Transcription," In Proc. ICASSP 2014.
\bibitem{dtw} H. Sakoe and S. Chiba, "Dynamic Programming Algorithm Optimization for Spoken Word Recognition," IEEE Transactions On
Acoustics, Speech and Signal Processing, Vol. 26, No. 1, February 1978.
\bibitem{segdtw} A. Park and J. Glass, "Unsupervised Pattern Discovery in Speech," , IEEE Transactions On Audio, Speech And Language Processing, Vol. 16, No. 1, January 2008.
\bibitem{kwssegdtw} Z. Yaodong and J. Glass, "Unsupervised Spoken Keyword Spotting via Segmental DTW on Gaussian Posteriorgrams," In Proc. ASRU 2009.
\bibitem{spksegdtw} A. Park and J. Glass, "A Novel DTW-Based Distance Measure for Speaker Segmentation," In Proc. SLT 2006.
\bibitem{PLDA} P. Kenny, "Bayesian Speaker Verification with Heavy-Tailed Priors," In Proc. Odyssey 2010.
\bibitem{shixiong2016} S. Zhang, Z. Chen, Y. Zhao, J. Li and Y. Gong, "End-to-End Attention Based Text-Dependent Speaker Verification," In Proc. SLT 2016.
\bibitem{x-vectors} David Snyder, Daniel Garcia-Romero, Gregory Sell, Daniel Povey, Sanjeev Khudanpur, "X-Vectors: Robust DNN Embeddings for Speaker Recognition" in Proc. ICASSP 2018.

\end{thebibliography}
\end{document}